\documentclass[apj, iop]{emulateapj}
\usepackage{natbib}
\usepackage{hyperref}
\usepackage{colortbl}
\usepackage{color}
\newcommand{\msun}{\,{\rm M_\odot}}

\newcommand{\beq}{\begin{equation}}
\newcommand{\eeq}{\end{equation}}
\newcommand{\ba}{\begin{eqnarray}}
\newcommand{\ea}{\end{eqnarray}}
\def\spose#1{\hbox to 0pt{#1\hss}}
\newcommand{\lta}{\mathrel{\spose{\lower 3pt\hbox{$\mathchar'218$}}
      \raise 2.0pt\hbox{$\mathchar"13C$}}}
\newcommand{\gta}{\mathrel{\spose{\lower 3pt\hbox{$\mathchar"218$}}
      \raise 2.0pt\hbox{$\mathchar"13E$}}}
\newcommand{\dd}{\mathrm{d}}

\definecolor{grey}{rgb}{0.75,0.75,0.75}
\definecolor{Orange}{rgb}{1.0,0.5,0.15}
\definecolor{brown}{rgb}{0.7,0.25,0.0}
\definecolor{pink}{rgb}{1.0,0.5,0.5}
\definecolor{darkerred}{rgb}{0.8,0,0}
\definecolor{darkerblue}{rgb}{0,0,0.8}
\definecolor{Blue}{rgb}{0,0.08,0.65}
\definecolor{Red}{rgb}{0.65,0.08,0.05}
\definecolor{Green}{rgb}{0.15,0.45,0.25}

\newcommand{\lumunits}{erg~s$^{-1}$~}

\shorttitle{High-z $M_{\rm BH} - M_{\star}$ Relations}
\shortauthors{Volonteri \& Reines}

\begin{document}

\title{Inferences on the Relations Between Central Black Hole Mass and Total Galaxy Stellar Mass in the high-redshift Universe}

\author{Marta Volonteri}
\affil{Institut d'Astrophysique de Paris, Sorbonne Universit\`{e}s, UPMC Univ Paris 6 et CNRS, UMR 7095, 98 bis bd Arago, 75014 Paris, France}

\and

\author{Amy E. Reines\altaffilmark{1}}
\affil{National Optical Astronomy Observatory, 950 North Cherry Avenue, Tucson, AZ 85719, USA}

\altaffiltext{1}{Hubble Fellow}

\begin{abstract}
At the highest redshifts, $z>6$, several tens of luminous quasars have been detected. The search for fainter AGN, in deep X-ray surveys, has proven less successful, with few candidates to date. An extrapolation of the relationship between black hole (BH) and bulge mass would predict that the sample of $z>6$ galaxies host relatively massive BHs ($>10^6 \msun$), if one assumes that total stellar mass is a good proxy for bulge mass. At least a few of these BHs should be luminous enough to be detectable in the 4Ms CDFS.  The relation between BH and stellar mass defined by local moderate-luminosity AGN  in low-mass galaxies, however, has a normalization that is lower by approximately an order of magnitude compared to the BH-bulge mass relation. We explore how this scaling changes the interpretation of AGN in the high-$z$ Universe. Despite large uncertainties, driven by those in the stellar mass function, and in the extrapolation of local relations, one can explain the current non-detection of moderate-luminosity AGN in Lyman Break Galaxies if galaxies below $10^{11} \msun$ are characterized by the low-normalization scaling, and, even more so, if their Eddington ratio is also  typical of moderate-luminosity AGN rather than luminous quasars. AGN being missed by X-ray searches due to obscuration or instrinsic X-ray weakness also remain  a possibility.
\end{abstract}


\section{Introduction}

The frontier of high redshift galaxies and quasars has now reached a relatively large sample. Hundreds of Lyman Break Galaxies (LBGs) with colors consistent with $z>6$ have been detected in deep fields \citep[e.g.,][and references therein]{2015arXiv151105558F}, and tens of luminous quasars are known at $z>6$ \citep[e.g.,][and references therein]{2012RAA....12..865F}.  The population of fainter active galactic nuclei (AGN) is still elusive. Partly, current surveys are not deep enough to detect them directly, and, partly, X-ray stacking of LBGs has led to no signal detected \citep{2011ApJ...742L...8W,2012AA...537A..16F,2012ApJ...748...50C,2013ApJ...778..130T}. Searches for point sources in deep X-ray fields has also led to inconclusive results \citep{2015AA...578A..83G,2015MNRAS.448.3167W,2015arXiv151200510C}. 

The X-ray non-detections have been used to estimate an upper limit on the black hole (BH) mass density at $z > 6$  through an analog of Soltan's argument \citep{Soltan1982}, and on the luminosity a putative AGN can have in these galaxies \citep{Treister2011,2013ApJ...778..130T}. With some assumptions on the Eddington ratio, this can be translated into an upper limit on the BH mass. The apparent result is that, if LBGs host BHs, they are accreting at low rate, or are less massive than expected on the basis of extrapolations of the correlation between BH mass and bulge mass at $z=0$ \citep{MarconiHunt2003,Haring2004,2013ARAA..51..511K}. However, it is far from clear if high-redshift LBGs have well developed bulges.

\citet[][RV15 thereafter]{2015ApJ...813...82R} have studied the relation between BH mass and total stellar mass for nearby galaxies ($z < 0.055$), including both galaxies with quiescent and active BHs. For the latter, the BH mass estimate is based on reverberation mapping or single-epoch virial estimates, the same technique used at higher redshift. Likewise, their stellar mass measurements rely on mass-to-light ratios, as done on higher redshift samples.  Therefore they  adopted the same methods used for mass measurements at higher redshift, where detailed information on stellar kinematics and bulge properties is not available. They found that the relation between BH mass and total stellar mass for moderate-luminosity AGN, predominantly hosted by lower-mass galaxies, has a normalization that is approximately an order of magnitude lower than BH-bulge mass relations largely constrained at high mass.

In this paper we assess whether the lower normalization identified for the  low-mass galaxies, typically lacking strong bulges, can explain the lack of an X-ray detection in the stack of LBGs. We couple galaxy stellar mass functions (MFs) with BH-stellar mass relations, and estimate the redshift evolution of the BH mass density and MF. We also take a complementary approach of coupling AGN luminosity functions at $z=6$ with an empirical Eddington ratio distribution, derived from the high-luminosity end of the luminosity function, to determine the BHMF.

\section{Method}
Our approach resembles that taken by \cite{Shankar2004}, \cite{2009MNRAS.399.1988S}, \cite{2010AJ....140..546W} and \cite{2011AA...535A..87S}. We start by paraphrasing some text from a paper by \cite{2011AA...535A..87S}.

We adopt the following convention: MBH masses are given by $\mu=\log M_{\rm BH}$,  the stellar mass  is $s$, with $s=\log M_*$, and the luminosity $l=\log L_{\rm AGN}$. Given a galaxy MF, $\Phi_*(s)$, and a function $g(\mu \, |\, s)$ which gives the probability of finding a BH of mass $\mu$ in a galaxy of mass $s$, the BHMF becomes: 
\begin{equation}
 \Phi_{\rm MBH, GAL}(\mu)=  \int g(\mu \, |\, s)\, \Phi_*(s)\, \dd s. 
\end{equation}
The integral of the BHMF then gives the mass density in BHs. Similarly, the integral of the galaxy MF gives the stellar mass density.  

Based on the empirical correlation between $\mu$ and $s$:  $\mu= \gamma + \alpha s$, with log-normal intrinsic scatter $\sigma$, i.e.
\begin{equation}
 g(\mu \, |\, s) = \frac{1}{\sqrt{2 \pi} \sigma} \exp \left\lbrace - \frac{(\mu-\gamma-\alpha s)^2}{2 \sigma^2} \right\rbrace .  \label{eq:gms}
\end{equation} 

 A similar approach links the AGN luminosity function, $\Phi_{\rm AGN}$, to the BHMF,  through $f(\lambda)$, the probability distribution of the logarithmic Eddington ratio $\lambda$, recalling that $l=38.11+\lambda+\mu$, and a duty cycle, ${\cal D}$:

\begin{equation}
\Phi_{\rm AGN}(\l)=  \int {\cal D} f(\lambda)  \Phi_{\rm MBH,AGN}(\mu) \dd l.
\end{equation}  

We consider here the $\Phi_{\rm MBH,AGN}(\mu)$ at $z=6$ derived by \cite{2010AJ....140..546W}, starting from the quasar luminosity function by \cite{Willott2010}, with $f(\lambda)$, fitted on the sample of  $z\sim6$ quasars with estimated BH mass, described by a lognormal distribution with $\bar{\lambda}=\log(0.6)$,  $\sigma=0.3$,  and ${\cal D}=0.75$.  

Additionally, a fraction of AGN are obscured, and they are missed by observations. We include a luminosity dependent correction for obscuration based on \cite{2014ApJ...786..104U}. Note that \cite{2014ApJ...786..104U} limit their redshift evolution to $z\sim2$. They found that the fraction of obscured quasars increases with redshift, but, conservatively, we keep the $z=2$ value even at higher redshift.

\begin{figure}[!t]
\hspace{-.4cm}
\includegraphics[width=3.5in]{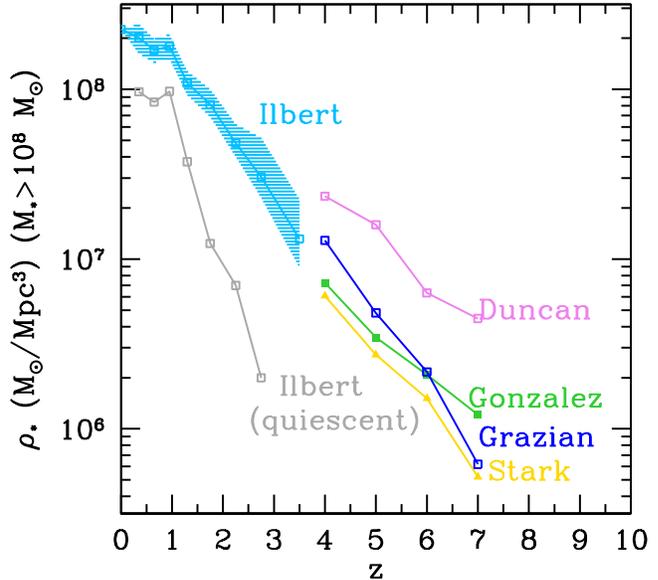}
\caption{\footnotesize  Comparison between the stellar mass density obtained by integration of different galaxy MFs for $M_*>10^8 \msun$.  In this work we adopt the MF by \cite{2013AA...556A..55I} at $z\leqslant 3.5$, and the MF by \cite{2015AA...575A..96G} at $z \geqslant 4$. However, it is important to keep in mind the spread in the observational determination as an important source of uncertainty in our results. 
\label{fig:rhogal}}
\end{figure}

\subsection{Galaxy mass functions}
Several different measurements and analytical fits to the galaxy stellar MF can be found in the literature. Many of them are summarized in \cite{2013ApJ...770...57B} and  \cite{2014ARAA..52..415M}, where differences and uncertainties are discussed (see Fig.11 in Madau \& Dickinson 2014). We will further discuss this in section 3.

We start from the galaxy MF of \cite{2013AA...556A..55I}. We use their best fit parameters for the full sample, and the fit for ``quiescent" galaxies as a proxy for elliptical galaxies. At $z>4$ we consider four galaxy MFs: \cite{2011ApJ...735L..34G}, plus the correction for nebular lines proposed by \cite{2013ApJ...763..129S}, \cite{2014MNRAS.444.2960D} and \cite{2015AA...575A..96G}, all converted to a Chabrier initial MF for consistency with RV15. The stellar mass density for the various MFs obtained by integration for stellar masses $>10^8 \msun$ is shown in Fig.~\ref{fig:rhogal}. In the following we will use as a reference the MF by \cite{2015AA...575A..96G} as ``middle ground", and discuss how results change using other MFs.

\subsection{BH-stellar mass relationships}
We adopt three different functional forms for the scaling between BH mass and galaxy stellar mass. The first is a simple linear scaling, so that the BH mass is $2\times 10^{-3}$ the stellar mass:
 \begin{equation}
 \mu=s-2.7,  
 \label{vanilla}
 \end{equation}
as often done in the literature, by extrapolating the BH-bulge mass relation of \cite{MarconiHunt2003} and \cite{Haring2004}. This is our ``vanilla" model.

 We also include the two total stellar mass relationships found by RV15 for ellipticals and bulges, typically with high stellar masses: 
\begin{equation}
\mu=1.40s-6.45
\label{fitQ}
\end{equation}
and for moderate-luminosity AGNs, typically in lower-mass host galaxies:
\begin{equation}
\mu=1.05s-4.10,
\label{fitAGN}
\end{equation}

\noindent ``HighMass" and ``LowMass" fits hereafter. Both these relationships have an intrinsic scatter $\sim 0.5$ dex.  In what follows we will adopt a scatter of 0.5 dex for all scalings as a reference and then discuss the effect of a tighter or broader scatter. We perform a Monte Carlo experiment with 50,000 draws for each BH or galaxy mass unless otherwise stated. 

\begin{figure}[!t]
\hspace{-.4cm}
\includegraphics[width=3.5in]{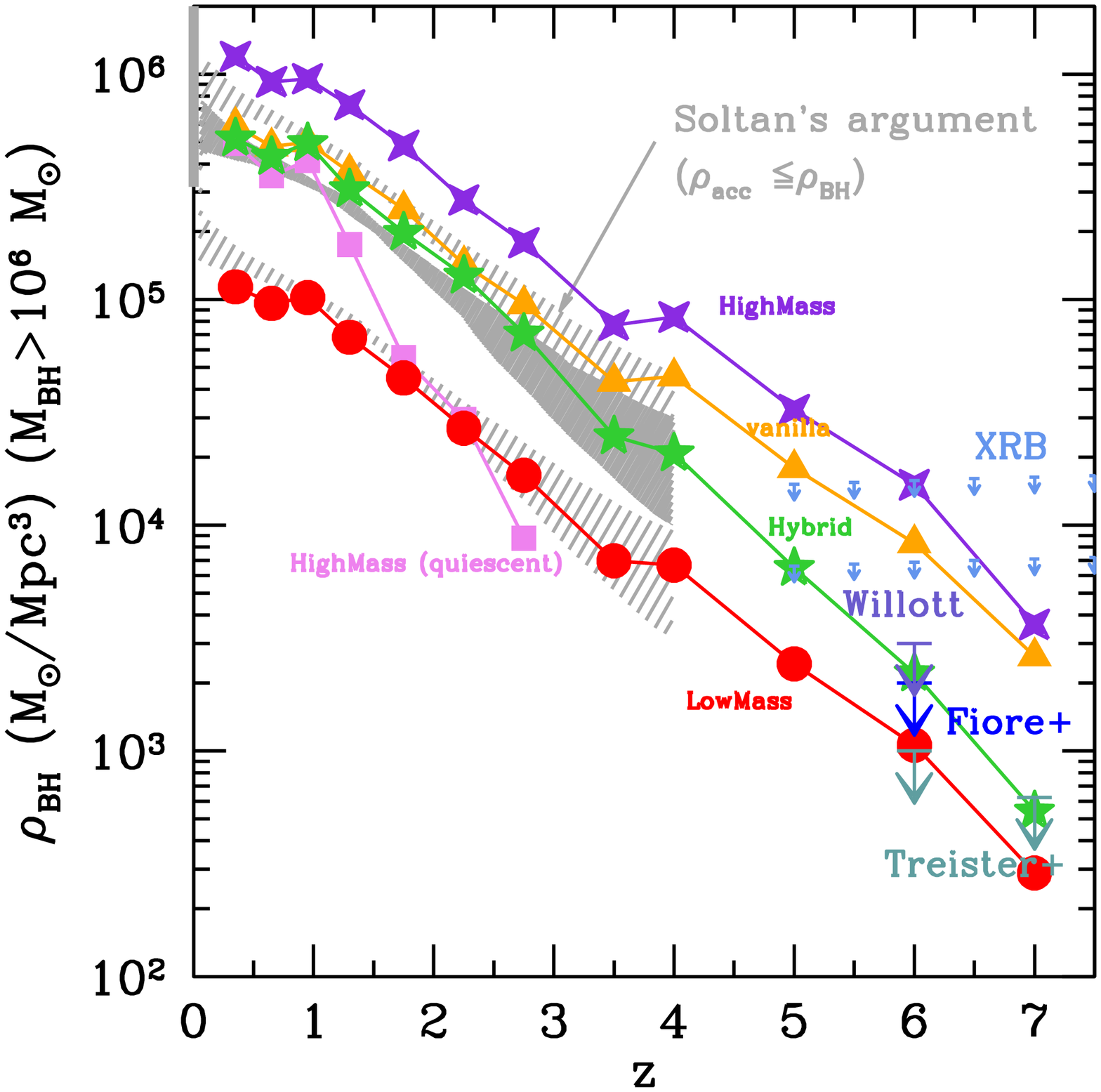}
\caption{\footnotesize  Mass density in BHs obtained by convolving the galaxy MF with different scalings of BH versus stellar mass. In all cases we assume a gaussian scatter of 0.5 dex, the MF of \cite{2013AA...556A..55I} at $z<4$ and the MF of \cite{2015AA...575A..96G} at $z>4$. Red circles: LowMass fit. Purple diamonds: HighMass fit. Green stars: LowMass fit below $10^{11} \msun$ and the HighMass fit above.  Orange triangles: fixed BH-stellar mass ratio of  $2\times 10^{-3}$, as often done in the literature, by extrapolating the BH-bulge mass relation of Marconi \& Hunt (2003) and Haring \& Rix (2004). Pink squares: HighMass fit and MF of \cite{2013AA...556A..55I} for quiescent galaxies only.  In all cases we include all BHs, quiescent and active. Grey hatched region: Soltan's argument.  Vertical grey line:  $z=0$ BH mass density. The limits at $z>6$ are derived from searches for AGN in stacked high-z galaxies or from the integrated X-ray background. These limits do not include Compton Thick AGN and require a BH to be active at some level, typically $>10^{42}-10^{43}$ \lumunits in the soft or hard X-ray band.
\label{fig:rhobh}}
\end{figure}

\section{Results}
\subsection{Evolution of BH mass density}
We start by looking at an integral quantity $\rho_{BH}$, the BH mass density versus redshift, integrating $\Phi_{\rm MBH, GAL}$ from $\mu=5$ to $\mu=9$. For reference, at $z=0$ we show the mass density obtained by \cite{2013CQGra..30x4001S}.   At $z>0$, the main constraints come from Soltan's argument, where  the AGN luminosity function  is integrated over time, from $t_{\rm max}$ to $t(z)$, and rescaled by a (fixed) radiative efficiency, $\epsilon$, to obtain the density of  mass accreted on BHs as a function of redshift:

\begin{equation}
\rho_{\rm BH,acc}(z)= \frac{1-\epsilon}{\epsilon c^2}\int_{t_{\rm max}}^{t(z)} dt \int dL\,L\,\Phi_{\rm AGN}(L,t).
\end{equation}

We adopt as a reference the estimate by \cite{2016LNP...905..101M} at $z<4$, including contributions of unobscured AGN, Compton-thin and Compton-thick  AGN, and $\epsilon=0.1$, and show also the cases with $\epsilon=0.06$ and $\epsilon=0.3$. At $z>6$ we report all the current upper limits, derived either on deep X-ray observations \citep{2011ApJ...742L...8W,2012ApJ...748...50C,2012AA...537A..16F,2013ApJ...778..130T} or from the integrated X-ray background \citep{2012AA...545L...6S}. These upper limits do not include Compton Thick AGN, so that in reality there may be a fraction of BHs not accounted for. We also stress that Soltan's argument estimates the mass density accreted in luminous phases throughout cosmic time up to $z$. The total mass density can be higher when accounting for non-radiative BH growth, e.g. via mergers, radiatively inefficient accretion or heavily obscured accretion episodes, and when including inactive BHs. The integral of $\Phi_{\rm MBH, GAL}$ instead provides the total mass density in BHs, irrespective of the luminosity.

In Fig.~\ref{fig:rhobh} we summarize the main results on the redshift evolution of the BH mass density. At $z<1$, there is a general consensus: taking the full MF of  \cite{2013AA...556A..55I}, and assuming the vanilla fit, or including quiescent galaxies only and fit HighMass give similar results. The reason is that, while the mass in galaxies locked in quiescent galaxies is about half of the total stellar mass density, the BH mass locked in elliptical galaxies dominates the full population because BHs represent a higher fraction of their stellar mass. Using the LowMass fit only, instead, leads to an underestimate of the total BH mass density.

Results become more interesting at higher redshift. Firstly, the fraction of stellar mass in quiescent galaxies drops significantly. Therefore, even considering that BHs represent a larger fraction of the stellar mass in ellipticals, the global contribution to the BH mass density falls. Therefore, if BHs require a bulge component, BHs represent a higher fraction of the stellar mass of the bulge at increasing redshift. Secondly, for the full population, the mass density in BHs is always above the limits imposed by lack of X-ray detections in stacking of high-z galaxies, except for the LowMass fit , i.e., for the other fits to hold, X-ray limits imply most of the BH mass density was not accreted in a luminous phase.

Increasing the scatter only increases the BH mass density \citep{Lauer2007b,2009MNRAS.399.1988S,2011MNRAS.417.2085V}. Even reducing the scatter to zero, however, the vanilla or HighMass fits overestimate $\rho_{\rm BH,acc}$ given by the observational constraints at high-$z$.  BHs represent a smaller fraction of the stellar mass of the galaxy at higher redshift, and/or local moderate-luminosity AGN are good proxies for the BH-to-host relationship at high-$z$. A combination of the LowMass and HighMass fit (``hybrid"), using $s=11$ as dividing line (RV15), provides a reasonable evolution of the mass density at all redshifts, with only a slight tension with most upper limits at $z>6$. For the same $\mu-s$ relation of Eq.~\ref{fitAGN}, the uncertainty given by the unknowns in the MF amount to $\sim$1 dex, with the MF by \cite{2014MNRAS.444.2960D} requiring the strongest (negative) evolution in the $\mu-s$ relation to accommodate observational upper limits. 

In summary, the choice of the scaling relation has clear consequences for the derived BH mass density. At low redshift these are less marked, since massive galaxies contribute significantly to the mass density. At high redshift, since massive galaxies are largely absent, the contribution from low mass galaxies is more important.

\begin{figure}[!t]
\hspace{-.4cm}
\includegraphics[width=3.5in]{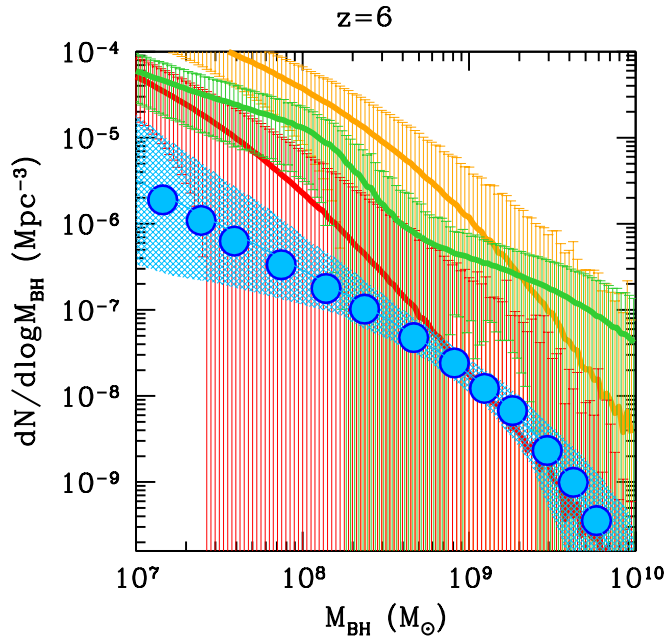}
\caption{\footnotesize  BHMF at $z=6$ obtained from either the AGN LF or the galaxy MF. Cyan dots and hatched region: $\Phi_{\rm MBH,AGN}$ from   \cite{2010AJ....140..546W}.  Red curve: LowMass fit. Light orange curve: vanilla fit.  Green curve:  LowMass fit below $10^{11} \msun$ and  HighMass fit above (hybrid fit). Scatter of 0.5 dex in each case. 
\label{fig:MF}}
\end{figure}

\subsection{Connection to the quasar population}

We focus here on what the scaling relations imply for the $z\sim 6$ luminous quasars, at the high-mass end of the BHMF, $\mu>8$. In Fig.~\ref{fig:MF} we compare $\Phi_{\rm MBH,GAL}$ to $\Phi_{\rm MBH,AGN}$ at $z=6$. With the vanilla and hybrid fits, $\Phi_{\rm MBH,GAL}>\Phi_{\rm MBH,AGN}$ \citep[see also the discussion in][]{2010AJ....140..546W,2011MNRAS.417.2085V}, requiring, e.g., a lower duty cycle or occupation fraction.  For the LowMass fit,  $\Phi_{\rm MBH,GAL}$ is in good agreement with $\Phi_{\rm MBH,AGN}$ at $\mu>9$. 

The masses of BHs powering the most luminous quasars, however, are estimated to be above the $z=0$ scaling \citep[assuming that the total dynamical mass corresponds to the stellar mass,][]{2013ApJ...773...44W}. To mimic their luminosity/flux limit, we associate a luminosity to BHs in $\Phi_{\rm MBH,GAL}$ through $f(\lambda)$, adopting the functional form and parameters given in section~2. If, for the LowMass fit, we select only BHs with bolometric luminosity $>10^{46}$ \lumunits, similar to currently-detected $z\sim 6$ quasars, this subset of the population, at $s<12$, is described by an {\it apparent} scaling between BH mass and galaxy stellar mass:

\begin{equation}
\mu=0.40s+3.87,
\label{fitbright}
\end{equation}

\noindent shallower and with a higher normalization than the scaling describing the full underlying population. This is a consequence of selection effects \citep{Lauer2007b,2011MNRAS.417.2085V}: at relatively low galaxy mass, only BHs  above the mean of the {\it intrinsic} scaling can reach very high luminosity.  BHs powering luminous quasars are more likely to lie above the intrinsic relation, which is recovered lowering the luminosity threshold.

\begin{figure}[!t]
\hspace{-.4cm}
\includegraphics[width=3.5in]{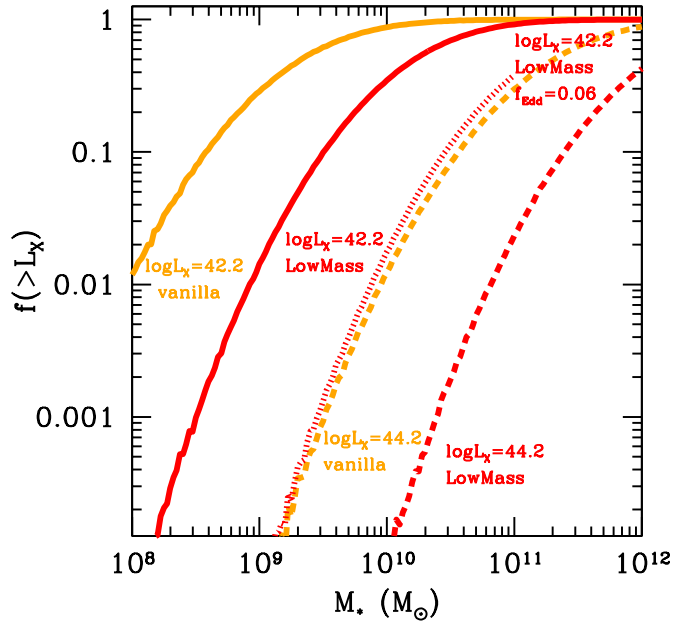}
\caption{\footnotesize  Fraction of galaxies at $z=6$ hosting an AGN above a given luminosity, marked in the figure, for the LowMass and vanilla fits. Solid and dashed curves:  $\bar{\lambda}=\log(0.6)$. Dotted curve: $\bar{\lambda}=\log(0.06)$, more typical of ``normal" AGN (this corresponds to the hybrid fit below $10^{11} \msun$).}
\label{fig:lum}
\end{figure}

\subsection{Implications for detecting AGN in LBGs}
X-ray stacking gives more direct upper limits on the luminosity of a putative AGN in LBGs, with typical stellar masses of $\sim 10^9 \msun$. According to Treister et al. (2013) at $z=6$ the luminosity in the hard X-ray band is $<1.6\times10^{42}$ \lumunits. We show in Fig.~\ref{fig:lum} the fraction of galaxies hosting an AGN  detectable above a given X-ray luminosity as a function of galaxy stellar mass, where we convert from bolometric luminosity to hard X-ray using the bolometric corrections of \cite{Marconi2004}.  We adopt again $\bar{\lambda}=\log(0.6)$, $ {\cal D}=0.75$ and a correction for obscuration. This Eddington ratio, however, was estimated on luminous quasars, and it is higher than the typical value for ``normal" AGN. The same applies to the duty cycle \citep[e.g.,][]{2010AA...516A..87S}. The absorbed fraction is also very conservative. We also assume that all galaxies host a BH. While today it is not clear how many galaxies with $s\sim 9$ have BHs \citep{2013ApJ...775..116R}, a LBG with $s\sim 9$ represented a massive galaxy at $z\sim 6$, and it is expected that such massive galaxies have been seeded with a BH by that time \citep{Volonteri2010AARV}.

Statistically, the fraction of galaxies with mass $\sim 10^{9} \msun$ hosting an unobscured AGN with $L_X>1.6\times10^{42}$ \lumunits is only $\sim$0.01 using the LowMass fit. Treister et al. (2013) stack 223 galaxies, and find no detection. Therefore the predicted luminosities are only slightly higher than the upper limit in the stack.   
If we select only BHs above this luminosity threshold, we can convert the mass function into an expected number of AGN in the 4Ms CDFS, covering about $10^{-6}$ of the sky area.  Between $z=6$ and $z=7$ we expect $2.22^{+0.79}_{-1.75}$ AGN with $L_X>1.6\times10^{42}$ \lumunits for the LowMass fit.  The vanilla fit gives $4.18^{+0.59}_{-1.46}$.

The accretion properties derived from luminous quasars are significantly different than the local Seyferts defining the LowMass fit,  making the estimates above conservative. The median Eddington ratio for the local AGN sample is around a factor of 10 lower (using the median $L_{\rm bol}$ and $M_{\rm BH}$ from the RV15 sample). With the LowMass fit, assuming a mean Eddington ratio of 0.06 in the lognormal distribution of Eddington ratios for BHs in galaxies with $s<11$, the fraction of  AGN at a given luminosity decreases (Fig.~\ref{fig:lum}), and  between $z=6$ and $z=7$ we expect $0.15^{+0.60}_{-0.15}$  AGN with $L_X>1.6\times10^{42}$ \lumunits in the 4Ms CDFS. For reference, the vanilla fit predicts $1.86^{+0.71}_{-1.77}$ AGN.

These results are based on the galaxy MF by \cite{2015AA...575A..96G}. For the galaxy MF predicting the largest number of galaxies, thus the most difficult to reconcile with a low number of BHs and AGN, \cite{2014MNRAS.444.2960D}, at $L_X>1.6\times10^{42}$ \lumunits  we find $3.40^{+0.84}_{-3.40}$ sources in the 4Ms CDFS area for the LowMass fit; $5.58^{+0.66}_{2.33}$ for the vanilla fit; in all cases adopting $\bar{\lambda}=\log(0.6)$, making these upper limits. Assuming $\bar{\lambda}=\log(0.06)$ at $s<11$, the numbers decrease to $0.65^{+1.19}_{-0.65}$ and $3.06^{+0.80}_{-3.06}$.

\section{Conclusions}
In this paper, we have drawn inferences on high-redshift BHs and their relation to their hosts. We have tested whether the relation between BH and galaxy stellar mass found by RV15 for local AGN ($z < 0.055$), can explain the lack of an X-ray detection in the stack of LBGs, because of the low normalization with respect to the BH-bulge mass relation characterizing bulge-dominated quiescent galaxies. We convolve galaxy stellar MFs with BH-stellar mass relations, and estimate the redshift evolution of the BH mass density and the BHMF at $z=6$.  We stress the speculative nature of this paper. It is very hard to draw firm, robust conclusions given the uncertainties on the observables. Despite the uncertainties, we can highlight some trends, and explain the current non-detection of moderate-luminosity AGN in LBGs using scaling relations for BH masses and AGN luminosities derived on observational samples. The main results can be summarized as follows:

\begin{itemize}
\item The fraction of stellar mass in quiescent galaxies drops significantly with redshift. If BHs require a bulge component, the ratio between BH and bulge mass must evolve positively with increasing redshift, in the sense that BHs represent a higher fraction of the stellar mass of the bulge. 
\item The total mass density in BHs is always above the limits imposed by lack of X-ray detections in stacking of high-z galaxies, except for the LowMass fit. Local moderate-luminosity AGN are good proxies for high-$z$ galaxies, and/or BHs represent a smaller fraction of the total stellar mass of the galaxy at high-$z$.
\item Using the BH-stellar mass scaling derived from local AGN hosted by low-mass galaxies jointly with, very conservatively, the accretion properties derived only from luminous quasars \citep{2010AJ....140..546W} is close to explaining the paucity of AGN in LBGs.  Moderate-luminosity AGN have lower Eddington ratios than luminous quasars, which makes the scarcity of AGN in LBGs even more reasonable. 
\item  If the BH-stellar mass scaling at high-$ z$ corresponds to today's BH-bulge mass, the lack of AGN in LBGs favors lower Eddington ratios for their BHs.
\end{itemize}

We have shown that using the empirical scaling between BH and galaxy mass, determined on local AGN  hosted by relatively low-mass galaxies, can explain the few, if any, moderate-luminosity AGN at $z>6$. One possibility is also that such AGN are intrinsically X-ray weak \citep{2014ApJ...794...70L}, or that obscuration is more important than currently thought.  Treister et al. (2013) also suggest alternative possibilities for such a low space density derived from the X-ray observations, among them a low BH occupation fraction at these redshift, a low AGN duty cycle, and/or BH growth through mergers. 

Getting firmer constraints on the mass of the host galaxies of the current sample of luminous quasars, and pushing at the same time for detections of AGN, e.g., using alternative techniques such as line ratios in the ultraviolet \citep{2016MNRAS.456.3354F} on the existing sample of LBGs would greatly help in understanding the link between BHs and galaxies at early times.

\acknowledgements 
We thank the referee for a careful and constructive review.  Support for AER was provided by NASA through Hubble Fellowship grant HST-HF2-51347.001-A awarded by the Space Telescope Science Institute, which is operated by the Association of Universities for Research in Astronomy, Inc., for NASA, under contract NAS 5-26555.
MV acknowledges funding from the European Research Council under the European Community's Seventh Framework Programme (FP7/2007-2013 Grant Agreement no.\ 614199, project ``BLACK'').  



\begin{thebibliography}{}
\expandafter\ifx\csname natexlab\endcsname\relax\def\natexlab#1{#1}\fi

\bibitem[{{Behroozi} {et~al.}(2013){Behroozi}, {Wechsler}, \&
  {Conroy}}]{2013ApJ...770...57B}
{Behroozi}, P.~S., {Wechsler}, R.~H., \& {Conroy}, C. 2013, \apj, 770, 57

\bibitem[{{Cappelluti} {et~al.}(2015){Cappelluti}, {Comastri}, {Fontana},
  {Zamorani}, {Amorin}, {Castellano}, {Merlin}, {Santini}, {Elbaz},
  {Schreiber}, {Shu}, {Wang}, {Dunlop}, {Bourne}, {Bruce}, {Buitrago},
  {Micha{\l}owski}, {Derriere}, {Ferguson}, {Faber}, \&
  {Vito}}]{2015arXiv151200510C}
{Cappelluti}, N., {Comastri}, A., {Fontana}, A., {et~al.} 2015, ArXiv e-prints,
  arXiv:1512.00510

\bibitem[{{Cowie} {et~al.}(2012){Cowie}, {Barger}, \&
  {Hasinger}}]{2012ApJ...748...50C}
{Cowie}, L.~L., {Barger}, A.~J., \& {Hasinger}, G. 2012, \apj, 748, 50

\bibitem[{{Duncan} {et~al.}(2014){Duncan}, {Conselice}, {Mortlock}, {Hartley},
  {Guo}, {Ferguson}, {Dav{\'e}}, {Lu}, {Ownsworth}, {Ashby}, {Dekel},
  {Dickinson}, {Faber}, {Giavalisco}, {Grogin}, {Kocevski}, {Koekemoer},
  {Somerville}, \& {White}}]{2014MNRAS.444.2960D}
{Duncan}, K., {Conselice}, C.~J., {Mortlock}, A., {et~al.} 2014, \mnras, 444,
  2960

\bibitem[{{Fan}(2012)}]{2012RAA....12..865F}
{Fan}, X. 2012, Research in Astronomy and Astrophysics, 12, 865

\bibitem[Feltre et al.(2016)]{2016MNRAS.456.3354F} Feltre, A., Charlot, S., 
\& Gutkin, J.\ 2016, \mnras, 456, 3354 

\bibitem[{{Finkelstein}(2015)}]{2015arXiv151105558F}
{Finkelstein}, S.~L. 2015, ArXiv e-prints, arXiv:1511.05558

\bibitem[{{Fiore} {et~al.}(2012){Fiore}, {Puccetti}, {Grazian}, {Menci},
  {Shankar}, {Santini}, {Piconcelli}, {Koekemoer}, {Fontana}, {Boutsia},
  {Castellano}, {Lamastra}, {Malacaria}, {Feruglio}, {Mathur}, {Miller}, \&
  {Pannella}}]{2012AA...537A..16F}
{Fiore}, F., {Puccetti}, S., {Grazian}, A., {et~al.} 2012, \aap, 537, A16

\bibitem[{{Giallongo} {et~al.}(2015){Giallongo}, {Grazian}, {Fiore}, {Fontana},
  {Pentericci}, {Vanzella}, {Dickinson}, {Kocevski}, {Castellano}, {Cristiani},
  {Ferguson}, {Finkelstein}, {Grogin}, {Hathi}, {Koekemoer}, {Newman}, \&
  {Salvato}}]{2015AA...578A..83G}
{Giallongo}, E., {Grazian}, A., {Fiore}, F., {et~al.} 2015, \aap, 578, A83

\bibitem[{{Gonz{\'a}lez} {et~al.}(2011){Gonz{\'a}lez}, {Labb{\'e}}, {Bouwens},
  {Illingworth}, {Franx}, \& {Kriek}}]{2011ApJ...735L..34G}
{Gonz{\'a}lez}, V., {Labb{\'e}}, I., {Bouwens}, R.~J., {et~al.} 2011, \apjl,
  735, L34

\bibitem[{{Grazian} {et~al.}(2015){Grazian}, {Fontana}, {Santini}, {Dunlop},
  {Ferguson}, {Castellano}, {Amorin}, {Ashby}, {Barro}, {Behroozi}, {Boutsia},
  {Caputi}, {Chary}, {Dekel}, {Dickinson}, {Faber}, {Fazio}, {Finkelstein},
  {Galametz}, {Giallongo}, {Giavalisco}, {Grogin}, {Guo}, {Kocevski},
  {Koekemoer}, {Koo}, {Lee}, {Lu}, {Merlin}, {Mobasher}, {Nonino}, {Papovich},
  {Paris}, {Pentericci}, {Reddy}, {Renzini}, {Salmon}, {Salvato}, {Sommariva},
  {Song}, \& {Vanzella}}]{2015AA...575A..96G}
{Grazian}, A., {Fontana}, A., {Santini}, P., {et~al.} 2015, \aap, 575, A96

\bibitem[{{H{\"a}ring} \& {Rix}(2004)}]{Haring2004}
{H{\"a}ring}, N., \& {Rix}, H.-W. 2004, ApJL, 604, L89

\bibitem[{{Ilbert} {et~al.}(2013){Ilbert}, {McCracken}, {Le F{\`e}vre},
  {Capak}, {Dunlop}, {Karim}, {Renzini}, {Caputi}, {Boissier}, {Arnouts},
  {Aussel}, {Comparat}, {Guo}, {Hudelot}, {Kartaltepe}, {Kneib}, {Krogager},
  {Le Floc'h}, {Lilly}, {Mellier}, {Milvang-Jensen}, {Moutard}, {Onodera},
  {Richard}, {Salvato}, {Sanders}, {Scoville}, {Silverman}, {Taniguchi},
  {Tasca}, {Thomas}, {Toft}, {Tresse}, {Vergani}, {Wolk}, \&
  {Zirm}}]{2013AA...556A..55I}
{Ilbert}, O., {McCracken}, H.~J., {Le F{\`e}vre}, O., {et~al.} 2013, \aap, 556,
  A55

\bibitem[{{Kashikawa} {et~al.}(2015){Kashikawa}, {Ishizaki}, {Willott},
  {Onoue}, {Im}, {Furusawa}, {Toshikawa}, {Ishikawa}, {Niino}, {Shimasaku},
  {Ouchi}, \& {Hibon}}]{2015ApJ...798...28K}
{Kashikawa}, N., {Ishizaki}, Y., {Willott}, C.~J., {et~al.} 2015, \apj, 798, 28

\bibitem[{{Kormendy} \& {Ho}(2013)}]{2013ARAA..51..511K}
{Kormendy}, J., \& {Ho}, L.~C. 2013, \araa, 51, 511

\bibitem[{{Lauer} {et~al.}(2007){Lauer}, {Tremaine}, {Richstone}, \&
  {Faber}}]{Lauer2007b}
{Lauer}, T.~R., {Tremaine}, S., {Richstone}, D., \& {Faber}, S.~M. 2007, ApJ,
  670, 249

\bibitem[{{Luo} {et~al.}(2014){Luo}, {Brandt}, {Alexander}, {Stern}, {Teng},
  {Ar{\'e}valo}, {Bauer}, {Boggs}, {Christensen}, {Comastri}, {Craig},
  {Farrah}, {Gandhi}, {Hailey}, {Harrison}, {Koss}, {Ogle}, {Puccetti}, {Saez},
  {Scott}, {Walton}, \& {Zhang}}]{2014ApJ...794...70L}
{Luo}, B., {Brandt}, W.~N., {Alexander}, D.~M., {et~al.} 2014, \apj, 794, 70

\bibitem[{{Madau} \& {Dickinson}(2014)}]{2014ARAA..52..415M}
{Madau}, P., \& {Dickinson}, M. 2014, \araa, 52, 415

\bibitem[{{Marconi} \& {Hunt}(2003)}]{MarconiHunt2003}
{Marconi}, A., \& {Hunt}, L.~K. 2003, ApJL, 589, L21

\bibitem[{{Marconi} {et~al.}(2004){Marconi}, {Risaliti}, {Gilli}, {Hunt},
  {Maiolino}, \& {Salvati}}]{Marconi2004}
{Marconi}, A., {Risaliti}, G., {Gilli}, R., {et~al.} 2004, MNRAS, 351, 169

\bibitem[{{Merloni}(2016)}]{2016LNP...905..101M}
{Merloni}, A. 2016, in Lecture Notes in Physics, Berlin Springer Verlag, Vol.
  905, Lecture Notes in Physics, Berlin Springer Verlag, ed. F.~{Haardt},
  V.~{Gorini}, U.~{Moschella}, A.~{Treves}, \& M.~{Colpi}, 101

\bibitem[{{Reines} {et~al.}(2013){Reines}, {Greene}, \&
  {Geha}}]{2013ApJ...775..116R}
{Reines}, A.~E., {Greene}, J.~E., \& {Geha}, M. 2013, \apj, 775, 116

\bibitem[{{Reines} \& {Volonteri}(2015)}]{2015ApJ...813...82R}
{Reines}, A.~E., \& {Volonteri}, M. 2015, \apj, 813, 82

\bibitem[{{Salvaterra} {et~al.}(2012){Salvaterra}, {Haardt}, {Volonteri}, \&
  {Moretti}}]{2012AA...545L...6S}
{Salvaterra}, R., {Haardt}, F., {Volonteri}, M., \& {Moretti}, A. 2012, \aap,
  545, L6

\bibitem[{{Schulze} \& {Wisotzki}(2010)}]{2010AA...516A..87S}
{Schulze}, A., \& {Wisotzki}, L. 2010, \aap, 516, A87

\bibitem[{{Schulze} \& {Wisotzki}(2011)}]{2011AA...535A..87S}
---. 2011, \aap, 535, A87

\bibitem[{{Schulze} {et~al.}(2015){Schulze}, {Bongiorno}, {Gavignaud},
  {Schramm}, {Silverman}, {Merloni}, {Zamorani}, {Hirschmann}, {Mainieri},
  {Wisotzki}, {Shankar}, {Fiore}, {Koekemoer}, \&
  {Temporin}}]{2015MNRAS.447.2085S}
{Schulze}, A., {Bongiorno}, A., {Gavignaud}, I., {et~al.} 2015, \mnras, 447,
  2085

\bibitem[{{Shankar}(2013)}]{2013CQGra..30x4001S}
{Shankar}, F. 2013, Classical and Quantum Gravity, 30, 244001

\bibitem[{{Shankar} {et~al.}(2004){Shankar}, {Salucci}, {Granato}, {De Zotti},
  \& {Danese}}]{Shankar2004}
{Shankar}, F., {Salucci}, P., {Granato}, G.~L., {De Zotti}, G., \& {Danese}, L.
  2004, MNRAS, 354, 1020

\bibitem[{{Soltan}(1982)}]{Soltan1982}
{Soltan}, A. 1982, MNRAS, 200, 115

\bibitem[{{Somerville}(2009)}]{2009MNRAS.399.1988S}
{Somerville}, R.~S. 2009, \mnras, 399, 1988

\bibitem[{{Stark} {et~al.}(2013){Stark}, {Schenker}, {Ellis}, {Robertson},
  {McLure}, \& {Dunlop}}]{2013ApJ...763..129S}
{Stark}, D.~P., {Schenker}, M.~A., {Ellis}, R., {et~al.} 2013, \apj, 763, 129

\bibitem[{{Treister} {et~al.}(2013){Treister}, {Schawinski}, {Volonteri}, \&
  {Natarajan}}]{2013ApJ...778..130T}
{Treister}, E., {Schawinski}, K., {Volonteri}, M., \& {Natarajan}, P. 2013,
  \apj, 778, 130

\bibitem[{{Treister} {et~al.}(2011){Treister}, {Schawinski}, {Volonteri},
  {Natarajan}, \& {Gawiser}}]{Treister2011}
{Treister}, E., {Schawinski}, K., {Volonteri}, M., {Natarajan}, P., \&
  {Gawiser}, E. 2011, Nature, 474, 356

\bibitem[{{Ueda} {et~al.}(2014){Ueda}, {Akiyama}, {Hasinger}, {Miyaji}, \&
  {Watson}}]{2014ApJ...786..104U}
{Ueda}, Y., {Akiyama}, M., {Hasinger}, G., {Miyaji}, T., \& {Watson}, M.~G.
  2014, \apj, 786, 104

\bibitem[{{Volonteri} \& {Stark}(2011)}]{2011MNRAS.417.2085V}
{Volonteri}, M., \& {Stark}, D.~P. 2011, \mnras, 417, 2085

\bibitem[{{Volonteri}(2010)}]{Volonteri2010AARV}
{Volonteri}, M. 2010, \aapr, 18, 279

\bibitem[{{Wang} {et~al.}(2013){Wang}, {Wagg}, {Carilli}, {Walter}, {Lentati},
  {Fan}, {Riechers}, {Bertoldi}, {Narayanan}, {Strauss}, {Cox}, {Omont},
  {Menten}, {Knudsen}, {Neri}, \& {Jiang}}]{2013ApJ...773...44W}
{Wang}, R., {Wagg}, J., {Carilli}, C.~L., {et~al.} 2013, \apj, 773, 44

\bibitem[{{Weigel} {et~al.}(2015){Weigel}, {Schawinski}, {Treister}, {Urry},
  {Koss}, \& {Trakhtenbrot}}]{2015MNRAS.448.3167W}
{Weigel}, A.~K., {Schawinski}, K., {Treister}, E., {et~al.} 2015, \mnras, 448,
  3167

\bibitem[{{Willott}(2011)}]{2011ApJ...742L...8W}
{Willott}, C.~J. 2011, \apjl, 742, L8

\bibitem[{{Willott} {et~al.}(2010{\natexlab{a}}){Willott}, {Albert},
  {Arzoumanian}, {Bergeron}, {Crampton}, {Delorme}, {Hutchings}, {Omont},
  {Reyl{\'e}}, \& {Schade}}]{2010AJ....140..546W}
{Willott}, C.~J., {Albert}, L., {Arzoumanian}, D., {et~al.} 2010{\natexlab{a}},
  \aj, 140, 546

\bibitem[{{Willott} {et~al.}(2010{\natexlab{b}}){Willott}, {Delorme},
  {Reyl{\'e}}, {Albert}, {Bergeron}, {Crampton}, {Delfosse}, {Forveille},
  {Hutchings}, {McLure}, {Omont}, \& {Schade}}]{Willott2010}
{Willott}, C.~J., {Delorme}, P., {Reyl{\'e}}, C., {et~al.} 2010{\natexlab{b}},
  \aj, 139, 906

\end{thebibliography}
\end{document}